\documentclass[%
 reprint,
 amsmath,amssymb,
 aps,
]{revtex4-2}

\usepackage{graphicx}
\usepackage{dcolumn}
\usepackage{bm}
\usepackage{booktabs} 
\usepackage{tabularx} 
\usepackage{caption} 
\usepackage{amsmath} 
\usepackage{lipsum} 
\usepackage{float} 

\begin{document}

\preprint{APS/123-QED}

\title{Probing 3D Velocity Distributions Insights from a Vibrated Dual-Species Granular System}

\author{Rameez Farooq Shah}
 \altaffiliation{rmzshah@gmail.com}
\author{Syed Rashid Ahmad}%
 \email{srahmad@jmi.ac.in}
\affiliation{%
 Department of Physics, Jamia Millia Islamia (A Central University), New Delhi 110025, India
}%

\begin{abstract}
We explore the velocity distributions in a vibrated binary granular gas system, focusing on how these distributions are influenced by the coefficient of restitution (CoR) and the inelasticity of particle collisions. Through molecular dynamics simulations, we examine the system's behavior for a range of CoR values below unity $\epsilon = 0.80, 0.85, 0.90, \text{ and } 0.95$
 and track the evolution of velocity distributions as the system approaches a non-equilibrium steady state. Our findings reveal a clear departure from the classical Maxwell-Boltzmann distribution, with increasing deviations as CoR decreases, indicating non-equipartition of energy between the two particle types. This behavior underscores the intricate dynamics inherent in inelastic collisions and highlights the significance of particle inelasticity in determining the system’s velocity distributions. These results not only provide insight into non-equilibrium statistical mechanics but also carry implications for real-world applications where granular materials are subject to external vibrations.

\end{abstract}

\maketitle


\section{\label{sec:level1}INTRODUCTION}
The study of granular materials—macroscopic particles interacting through dissipative collisions—has garnered significant attention due to their relevance in various industries and natural phenomena. These materials exhibit properties akin to both fluids and solids, forming heaps and withstanding deformation like solids while also flowing through constrictions like liquids and behaving like gases when agitated \cite{rmp_behringer, rmp_kadanoff}. 
Granular materials are characterized by their constituent particles, typically larger than 1 $\mu$m and polydisperse in size and shape. Their macroscopic nature means they are not subject to thermal fluctuations. In theoretical and numerical studies, these particles are often modeled as spheres, needles, or cylinders \cite{rmp_tsimring, duran, ristow, nb_ktgg}. 
One of the most distinctive properties of granular materials is their dissipative interactions, which lead to kinetic energy loss or cooling, accompanied by a local parallelization of particle velocities. These interactions result in fascinating phenomena, such as size separation, clustering, pattern formation, and anomalous velocity statistics \cite{haff83, swinney9596, gz93, mcny9296, jjbrey9698, tpcvn9798, sl9899, ap0607, adsp1213}. 
The granular gas, representing dilute granular systems, serves as a paradigm for understanding gas-like properties where molecules dissipate energy through interactions. In the absence of external energy input, a granular gas loses kinetic energy due to inelastic collisions. Initially, the system may exhibit a homogeneous cooling state (HCS), but it evolves into an inhomogeneous cooling state (ICS) due to fluctuations in density and velocity fields \cite{haff83, ap0607, dp03}.
In experimental settings, energy loss is often compensated by input through various drive mechanisms, such as horizontal or vertical vibration. Under these conditions, the system settles into a nonequilibrium steady state \cite{swinney9596, ristow}. The study of velocity distributions in driven granular systems is of particular interest, as these distributions deviate from the Maxwell-Boltzmann distribution observed in elastic gases \cite{ap0607, adsp1213, pdsp2018}.
This paper focuses on analyzing velocity distributions in a vibrated binary granular gas system. By examining a two-component mixture of granular particles subject to external vibration, we aim to elucidate the complex dynamics and statistical properties that emerge from the interplay of dissipative collisions, energy input, and particle interactions in a binary system.
Using molecular dynamics simulations, we investigate the velocity statistics of this vibrated binary granular gas. We explore how factors such as particle size ratios, mass ratios, and vibration parameters influence the velocity distributions of each component and the system as a whole. Through this analysis, we seek to contribute to the broader understanding of nonequilibrium phenomena in driven granular systems and their implications for various industrial and natural processes.

We begin with the Boltzmann equation for a granular gas, which describes the evolution of the one-particle velocity distribution function $f(\mathbf{v}, t)$ \cite{nb_ktgg, Noije}:
\begin{equation}
\frac{\partial f}{\partial t} + \mathbf{v} \cdot \nabla f = J[f,f]
\end{equation}
where $J[f,f]$ is the collision integral. For inelastic hard spheres, this can be written as:
\begin{equation}
J[f,f] = \sigma^2 \int d\mathbf{v}_1 \int d\hat{\mathbf{k}} (\mathbf{v}_{12} \cdot \hat{\mathbf{k}}) 
[\chi^{-2} f(\mathbf{v}') f(\mathbf{v}_1') - f(\mathbf{v}) f(\mathbf{v}_1)]
\end{equation}
Here, $\sigma$ is the particle diameter, $\mathbf{v}_{12} = \mathbf{v}_1 - \mathbf{v}$, $\hat{\mathbf{k}}$ is the unit vector along the line of centers at collision, and $\chi = 1/\epsilon$ is the inverse of the coefficient of restitution.
In the absence of external forcing, the system evolves towards a homogeneous cooling state (HCS). The velocity distribution in the HCS can be approximated by:
\begin{equation}
f_{HCS}(\mathbf{v}, t) = n v_0(t)^{-d} \phi(\mathbf{c})
\end{equation}
where $n$ is the number density, $d$ is the dimensionality, $\mathbf{c} = \mathbf{v} / v_0(t)$ is the scaled velocity, and $v_0(t)$ is the thermal velocity given by:
\begin{equation}
v_0(t) = \sqrt{\frac{2T(t)}{m}}
\end{equation}
The temperature $T(t)$ decreases according to Haff's law:
\begin{equation}
\frac{dT}{dt} = -\zeta T
\end{equation}
where $\zeta$ is the cooling rate.
For a driven granular gas, we modify the Boltzmann equation to include a forcing term $F[f]$:
\begin{equation}
\frac{\partial f}{\partial t} + \mathbf{v} \cdot \nabla f = J[f,f] + F[f]
\end{equation}
In our vibrated system, the forcing term can be modeled as a diffusion process in velocity space:
\begin{equation}
F[f] = D \nabla_v^2 f
\end{equation}
where $D$ is the diffusion coefficient related to the vibration strength.
To account for the observed deviations from the Maxwell-Boltzmann distribution, we can expand the velocity distribution function using Sonine polynomials:
\begin{equation}
f(\mathbf{v}) = f_{MB}(\mathbf{v}) [1 + a_2 S_2(c^2) + a_3 S_3(c^2) + ...]
\end{equation}
where $f_{MB}(\mathbf{v})$ is the Maxwell-Boltzmann distribution, $S_n(x)$ are Sonine polynomials, and $a_n$ are the expansion coefficients. The first few Sonine polynomials are:
\begin{align}
S_0(x) &= 1 \\
S_1(x) &= -x + \frac{d}{2} \\
S_2(x) &= \frac{1}{2}(x^2 - (d+2)x + \frac{d(d+2)}{4})
\end{align}



For the high-energy tail of the distribution, we can consider a stretched exponential form:
\begin{equation}
f(c) \sim \exp(-\beta c^\alpha)
\end{equation}
where $\alpha$ and $\beta$ are fitting parameters. Taking the logarithm of both sides:
\begin{equation}
\log(f(c)) \sim -\beta c^\alpha
\end{equation}
This allows us to determine $\alpha$ and $\beta$ through linear regression of $\log(-\log(f(c)))$ vs. $\log(c)$.
The paper is organized as follows: Section I provides an introduction to granular materials and their unique properties, emphasizing their relevance in various natural and industrial processes.
Section I.A describes the model and simulation methodology employed for the vibrated binary granular gas system. This includes details on the simulation setup, particle interactions, and the evolution of the system under different coefficients of restitution.
Section II focuses on the vibration mechanism in the system and explains how energy is introduced via vibrations, which drive the system towards a non-equilibrium steady state.
Section III presents the results of the velocity distribution analysis, highlighting key findings such as deviations from the Maxwell-Boltzmann distribution and the role of inelastic collisions in shaping the system’s velocity statistics. The characterization of the velocity distribution function in terms of the coefficients of the Sonine polynomial expansion is also discussed in Sec.III.
In III.A we calculated Statistical Moments ,the kurtosis and skew-ness for each direction. 
Section IV summarizes the conclusions, discussing their broader implications for granular physics and potential applications in industrial processes involving vibrated granular materials.
\subsection{\label{sec:level2}NUMERICAL SIMULATION}
We investigate the velocity distributions in a vibrated binary granular gas system through molecular dynamics simulations. Our model and methodology are designed to capture the complex dynamics of a two-component granular gas subject to vibration, allowing us to analyze the resulting velocity statistics under various conditions.
The system comprises $N = 500{,}000$ particles of two types, confined in a 3D cubical box with periodic boundary conditions. The initial configuration is characterized by:
Number density: $n = 0.02$, Particle positions: Randomly assigned to ensure no core overlaps.
Initial velocities: Randomly chosen such that the total velocity $\sum \mathbf{v}_i = 0$.
Our simulation procedure consists of the following steps:
Initial Equilibration, The system is evolved to a time $\tau = 100$ with a coefficient of restitution $\epsilon= 1$ (elastic collisions) to reach a Maxwell-Boltzmann (MB) velocity distribution. This serves as the starting condition for further simulations.
Inelastic Evolution Subsequently, the system is evolved to $\tau = 1000$ for four different coefficients of restitution: $\epsilon = 0.95, 0.90, 0.85, \text{ and } 0.80$.
Collision Dynamics, We employ an event-driven algorithm to simulate particle collisions. When particles $i$ and $j$ collide, their post-collision velocities are given by:
\begin{equation}
    \mathbf{v}_i' = \mathbf{v}_i - \frac{1 + \epsilon}{2} \left[(\mathbf{v}_i \cdot \hat{r}_{ij}) - (\mathbf{v}_j \cdot \hat{r}_{ij}) \right] \hat{r}_{ij},
\end{equation}
where $0 < \epsilon < 1$ is the coefficient of restitution and $\hat{r}_{ij}$ is the unit vector connecting the centers of particles $i$ and $j$.
\section{Vibration Mechanism in the System}
We examine the mechanism of driving the granular system via \textit{vibration along the x-axis}. The particles gain energy through interactions induced by the vibration at each time step, denoted as $\Delta t$.
In this method, particles receive energy through random increments in their velocities along the x-axis during each time step. The velocity update for a particle $i$ is given by:
\begin{equation}
    \mathbf{v}_i(t + \Delta t) = \mathbf{v}_i(t) + h \Delta t \mathbf{f}_x(t),
\end{equation}
where $\mathbf{f}_x(t)$ is a random vector affecting only the x-component of the velocity, uniformly distributed in the range $[-1/2, 1/2]$, and $h$ is proportional to the vibration strength. 

\begin{figure*}[htbp] 
    \centering
    
    \begin{minipage}[b]{0.45\textwidth}
        \centering
        \includegraphics[width=\textwidth]{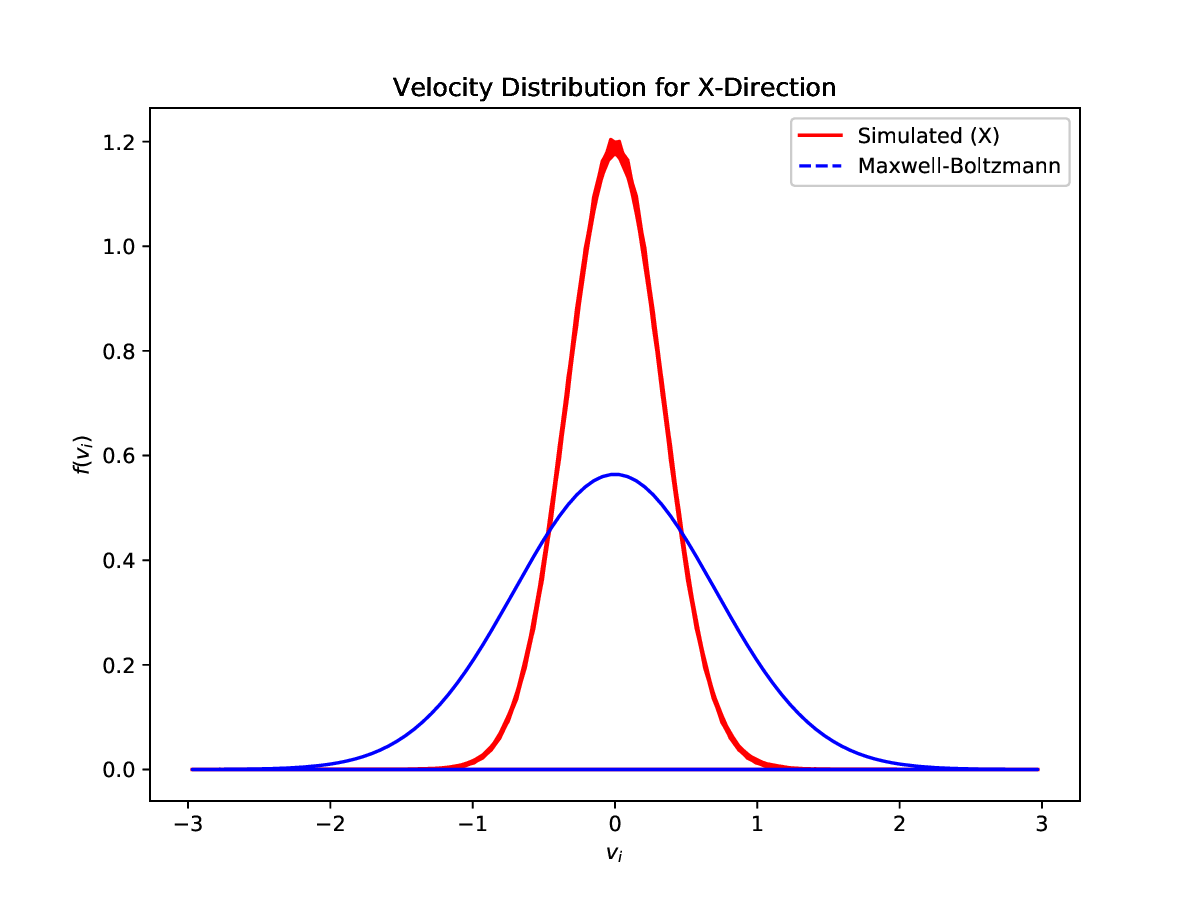}
        \caption*{For restitution coefficient $\epsilon = 0.80$.}
        \label{fig:velocity_distribution_X0.80}
    \end{minipage}
    \hfill
    \begin{minipage}[b]{0.45\textwidth}
        \centering
        \includegraphics[width=\textwidth]{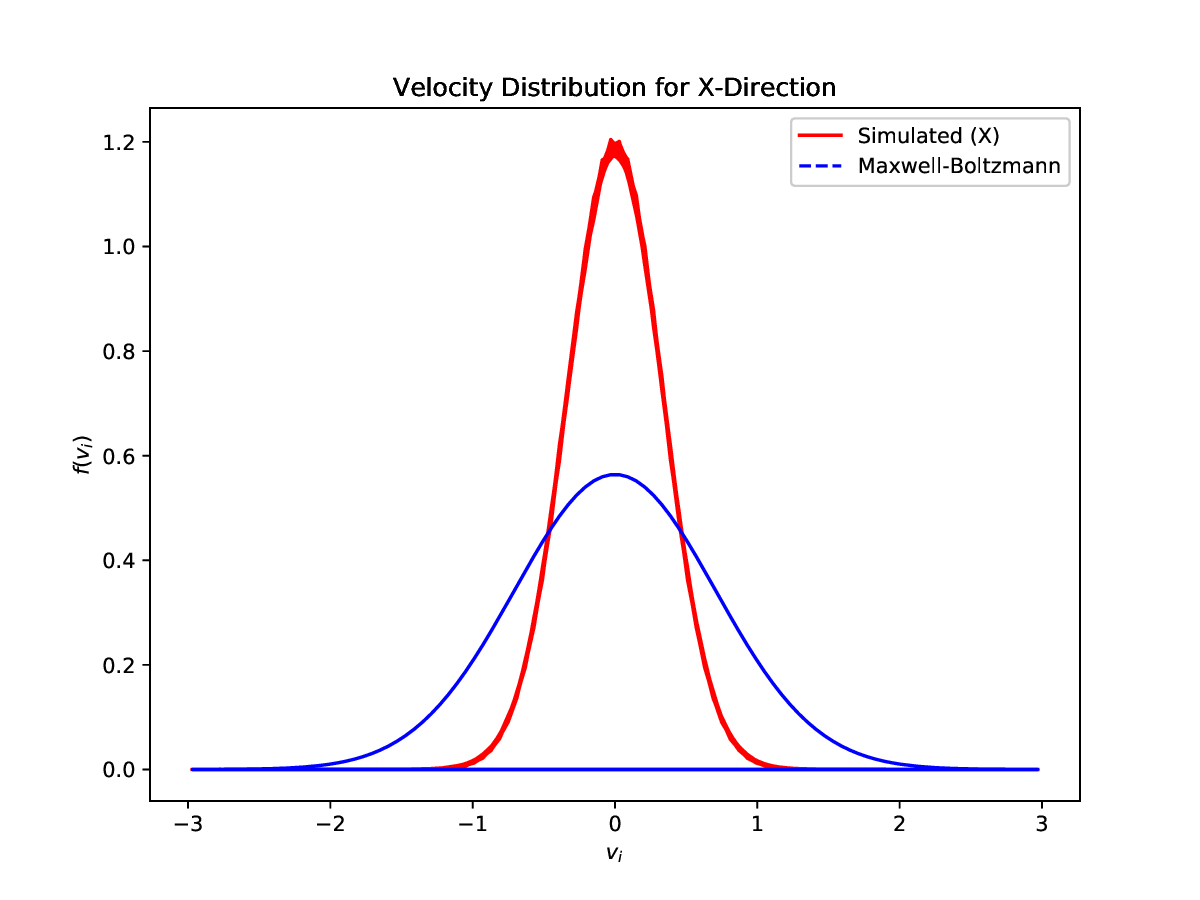}
        \caption*{For restitution coefficient $\epsilon = 0.85$.}
        \label{fig:velocity_distribution_X0.85}
    \end{minipage}

    \vskip\baselineskip
    
    \begin{minipage}[b]{0.45\textwidth}
        \centering
        \includegraphics[width=\textwidth]{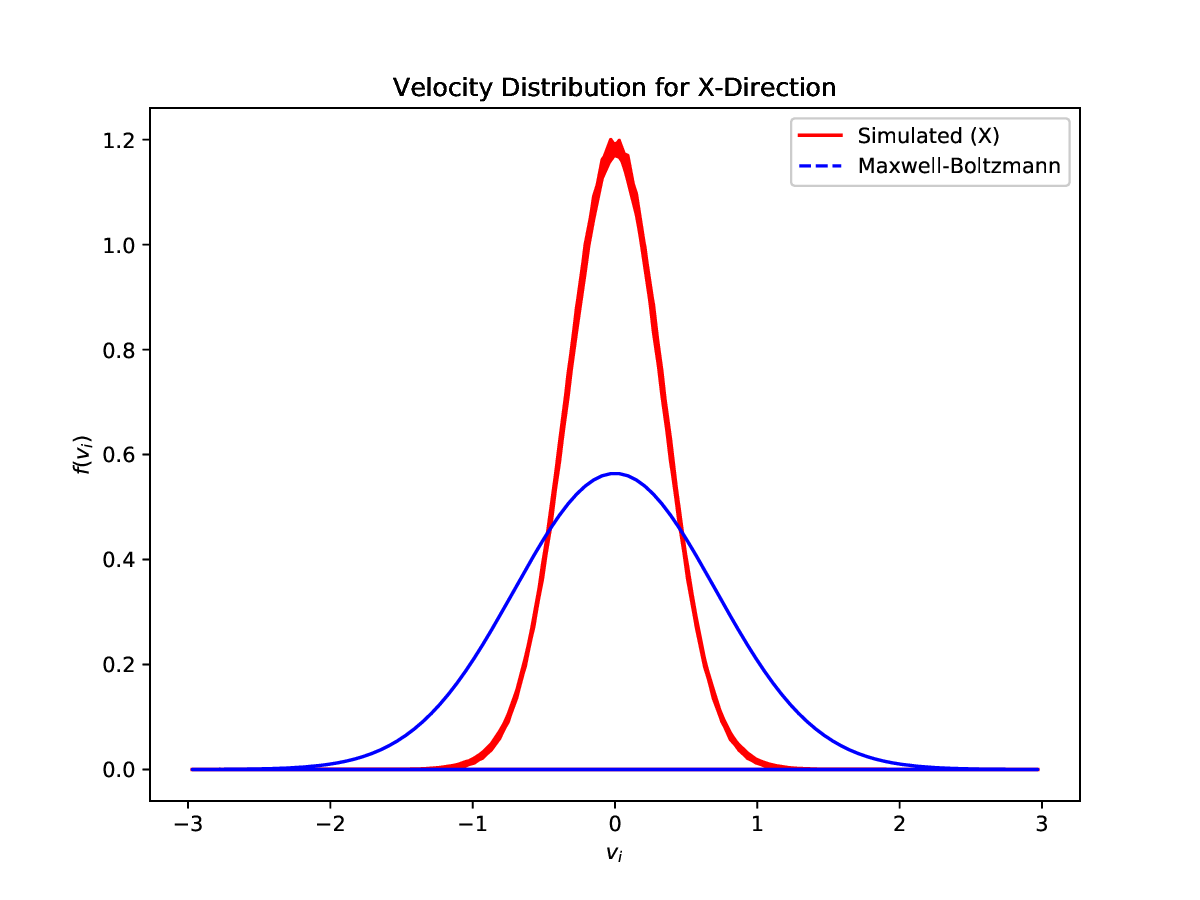}
        \caption*{For restitution coefficient $\epsilon = 0.90$.}
        \label{fig:velocity_distribution_X0.90}
    \end{minipage}
    \hfill
    \begin{minipage}[b]{0.45\textwidth}
        \centering
        \includegraphics[width=\textwidth]{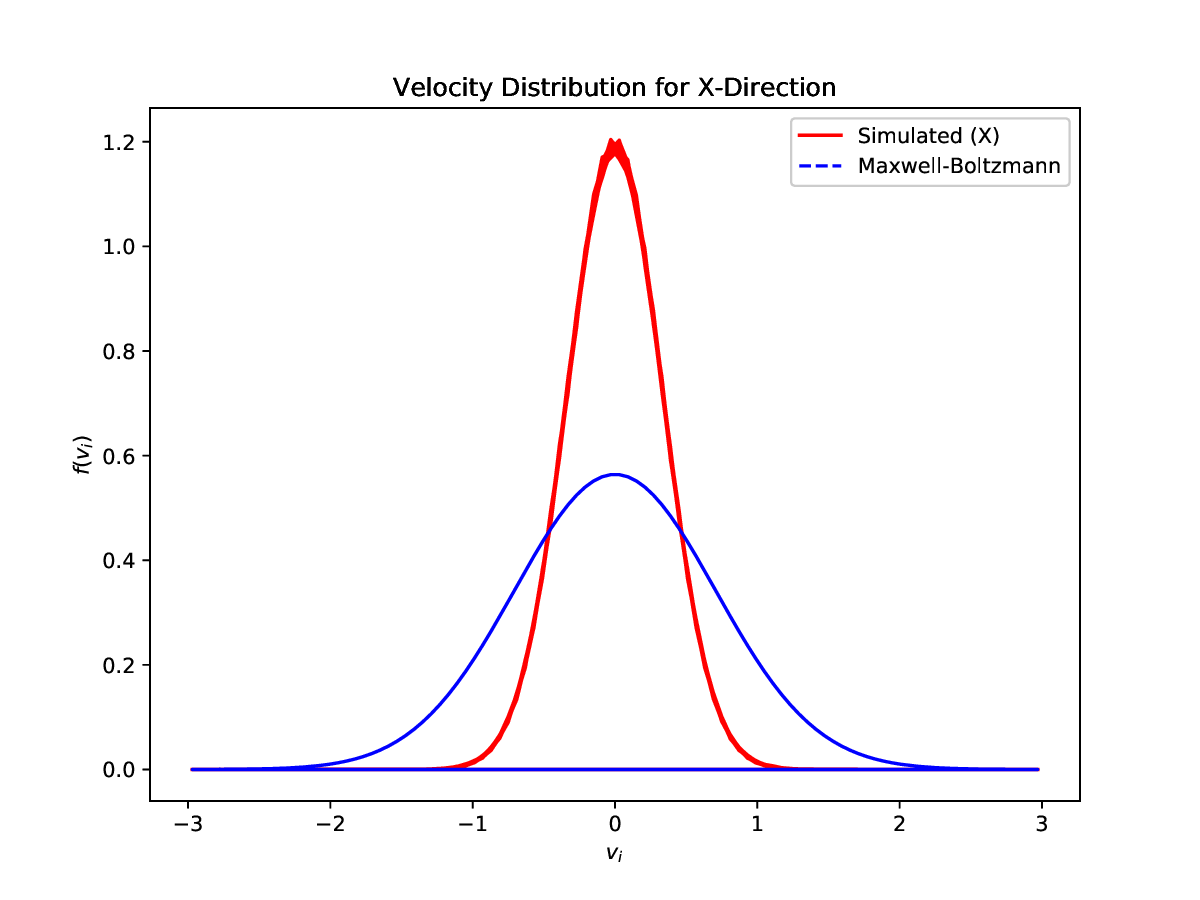}
        \caption*{For restitution coefficient $\epsilon = 0.95$.}
        \label{fig:velocity_distribution_X0.95}
    \end{minipage}

    \caption{ Velocity distributions and ratios for the Z-direction in a vibrated granular system. The ratio plots clearly show that our distribution has heavier tails than the MB distribution, with the ratio exceeding 1 for larger velocity magnitudes. This indicates an increased probability of finding particles with velocities much larger than the mean in our vibrated granular system.}
    \label{fig:velocity_distribution}
\end{figure*}
After applying the vibration, the system is transformed into the center-of-mass frame for analysis. Particles are confined to a cubical box with periodic boundary conditions to simulate bulk behavior. The time step $\Delta t$ is chosen such that, on average, fewer than one collision occurs per time step.
Unlike spatially homogeneous driving methods used in some experiments \cite{ref17}, where all particles experience uniform forcing, our vibration method introduces randomness, resulting in weak spatial and temporal correlations between neighboring particles.
At the start of the simulation, particles are uniformly distributed within the box. To facilitate effective interactions, they are assigned small, uniformly distributed velocities. The system is vibrated until it reaches a steady state before data collection begins. For the vibration mechanism, data are collected at every time step $\Delta t$.
\section{Velocity Distribution Analysis}
\begin{figure*}[htbp] 
    \centering
    
    \begin{minipage}[b]{0.45\textwidth}
        \centering
        \includegraphics[width=\textwidth]{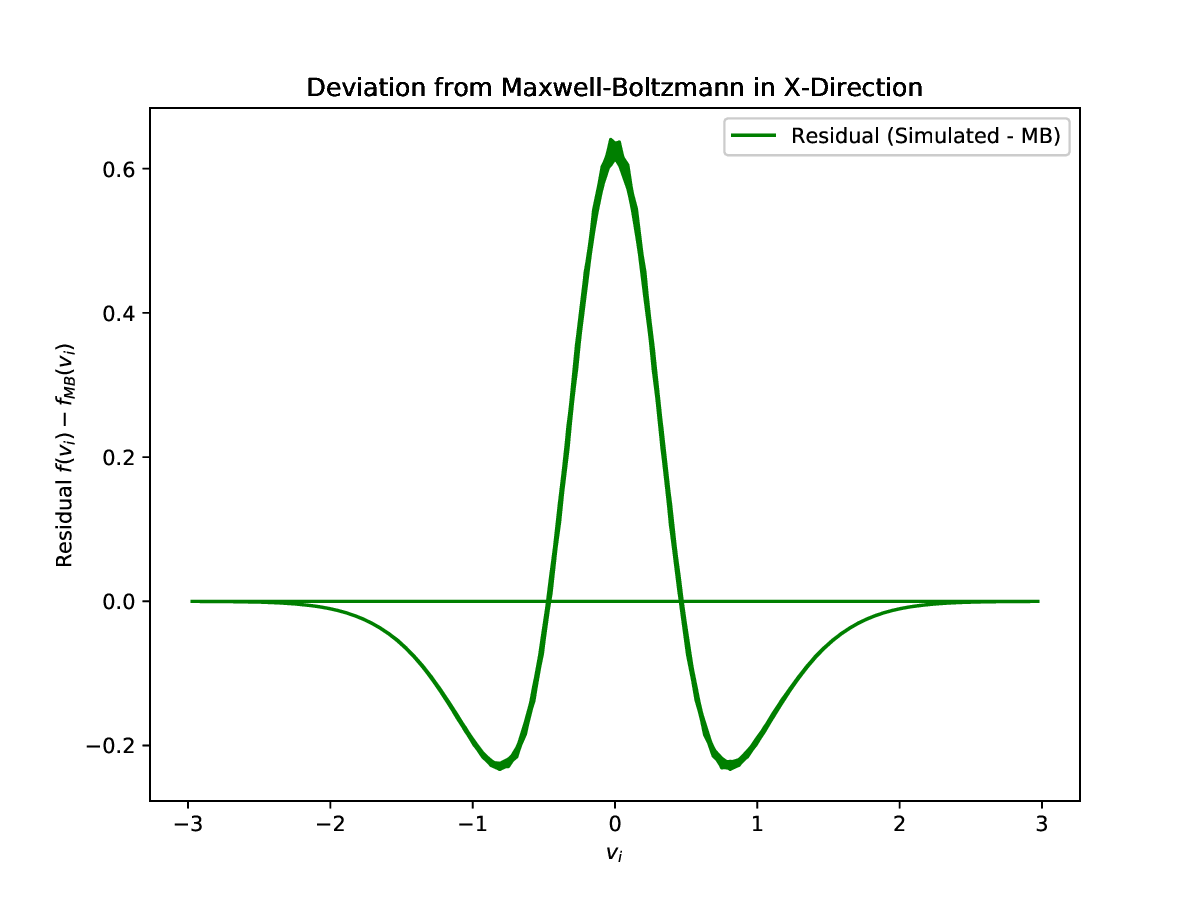}
        \caption*{Residuals (difference from Maxwell-Boltzmann) for the X-direction velocity distribution. For restitution coefficient $\epsilon = 0.80$.}
        \label{fig:residual_0.80}
    \end{minipage}
    \hfill
    \begin{minipage}[b]{0.45\textwidth}
        \centering
        \includegraphics[width=\textwidth]{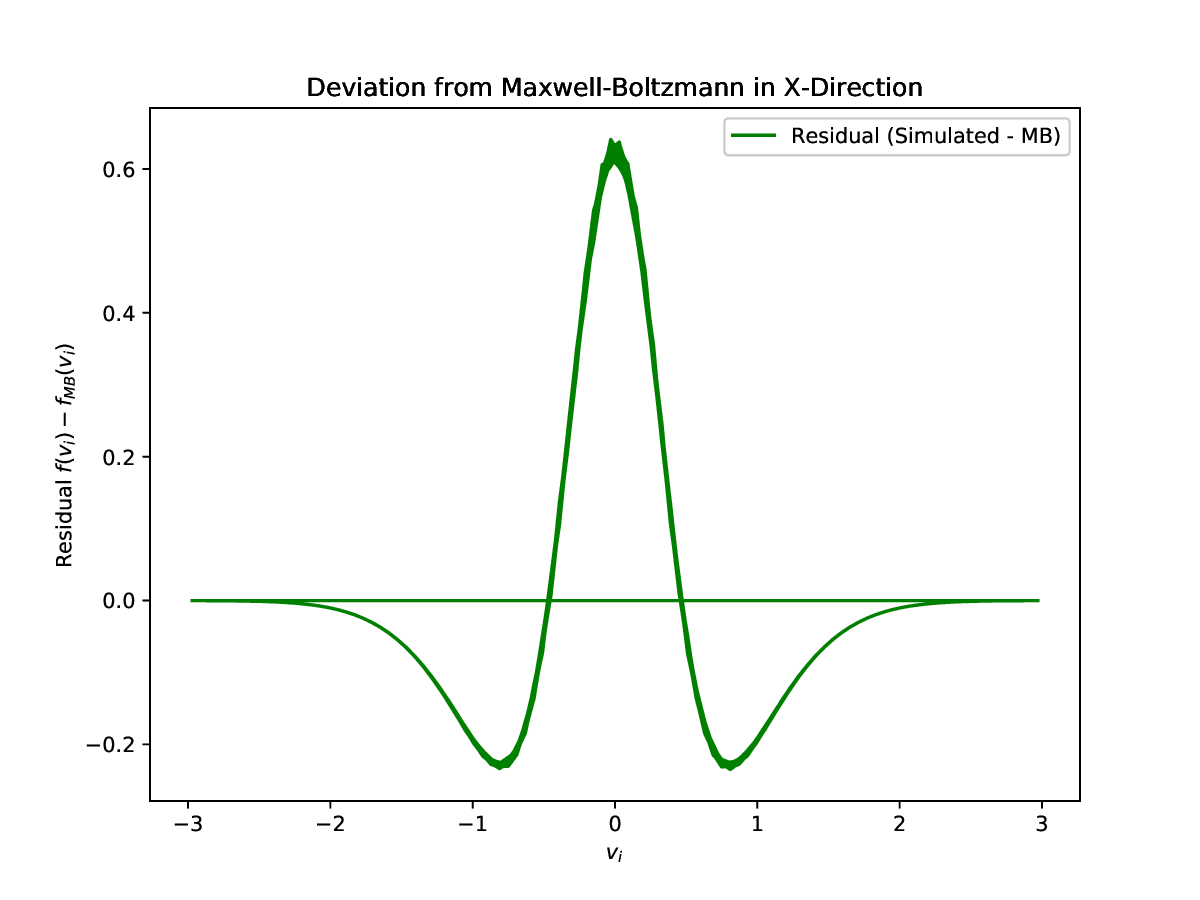}
        \caption*{Residuals (difference from Maxwell-Boltzmann) for the X-direction velocity distribution. For restitution coefficient $\epsilon = 0.85$.}
        \label{fig:residual_0.85}
    \end{minipage}

    \vskip\baselineskip
    
    \begin{minipage}[b]{0.45\textwidth}
        \centering
        \includegraphics[width=\textwidth]{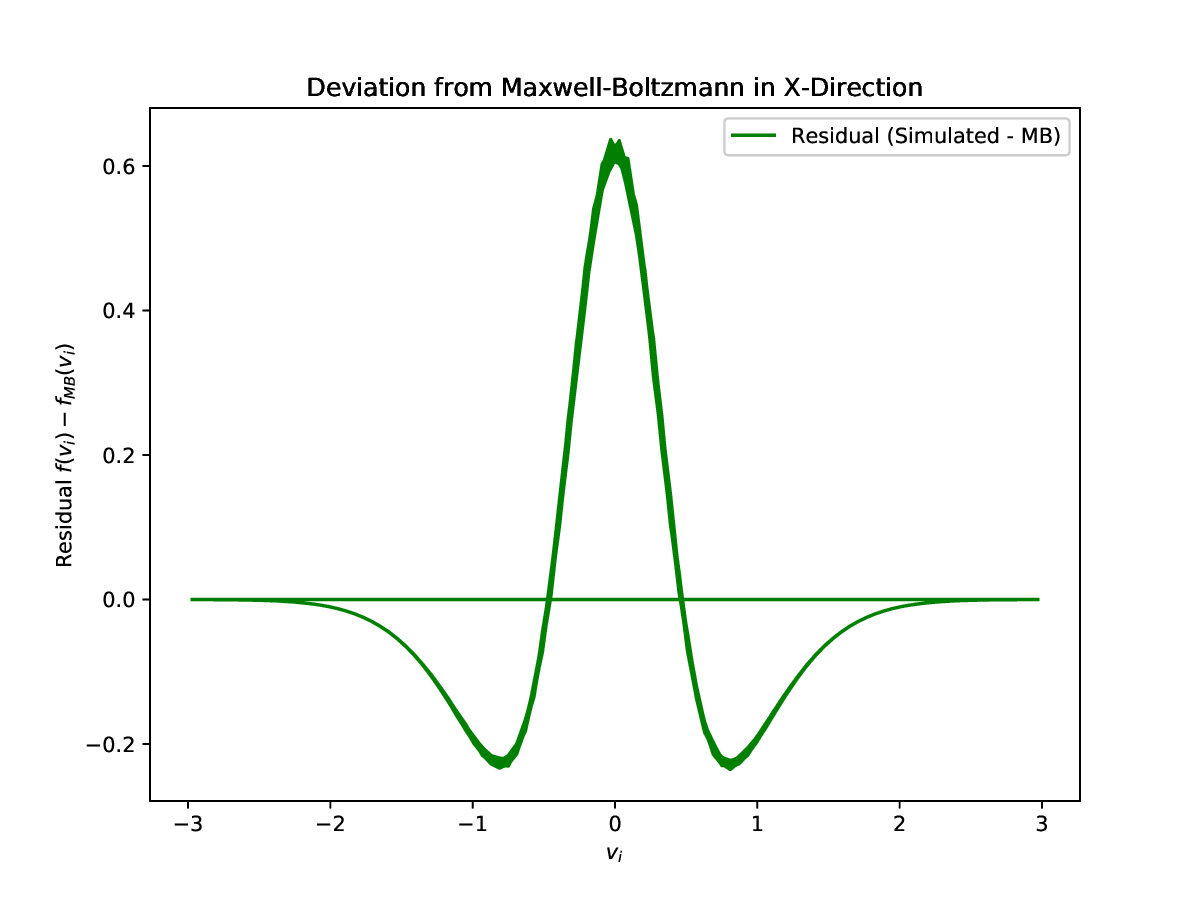}
        \caption*{Residuals (difference from Maxwell-Boltzmann) for the X-direction velocity distribution. For restitution coefficient $\epsilon = 0.90$.}
        \label{fig:residual_0.90}
    \end{minipage}
    \hfill
    \begin{minipage}[b]{0.45\textwidth}
        \centering
        \includegraphics[width=\textwidth]{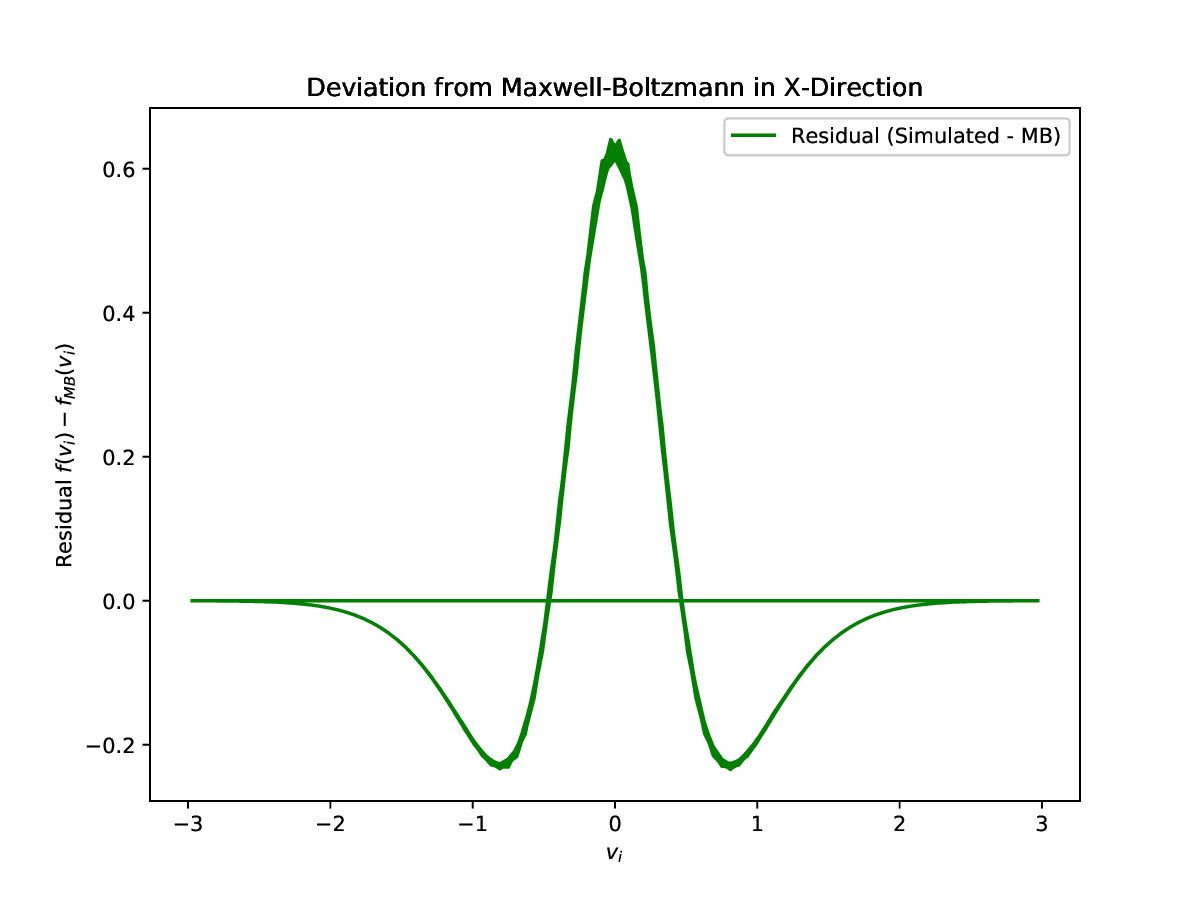}
        \caption*{Residuals (difference from Maxwell-Boltzmann) for the X-direction velocity distribution. For restitution coefficient $\epsilon = 0.95$.}
        \label{fig:residual_0.95}
    \end{minipage}

    \caption{ Residuals for X-direction velocity distributions. The residuals reveal systematic deviations from the MB distribution, with positive residuals in the tails indicating an excess of high-velocity particles compared to the equilibrium expectation.}
    \label{fig:residual_z}
\end{figure*}

To study the velocity distributions in our vibrated binary granular gas system, we begin by considering the standard Maxwell-Boltzmann (MB) distribution as a reference point. The MB distribution, applicable to equilibrium systems, is given by:
\begin{equation}
\label{eq:MB_VDF}
P_{\mathrm{MB}}(\vec{v})=\left(\frac{1}{\pi v_0^2}\right)^{d / 2} \exp \left(-\frac{\vec{v}^2}{v_0^2}\right), \quad v_0^2=\frac{2\left\langle\vec{v}^2\right\rangle}{d}
\end{equation}
where $d$ is the dimensionality of the system, and $v_0$ is the characteristic velocity related to the mean square velocity. 
The velocity distribution function evolves over time as a result of the cooling process, particularly in cases where the system is nearly elastic ($\epsilon \approx 1$). This function takes a scaling form that differs from the classical Maxwell-Boltzmann distribution \cite{gs95, vne98}:
\begin{equation}
P(\vec{v}, t) = \frac{1}{v_0^d(t)} G\left( \frac{\vec{v}}{v_0(t)} \right) \equiv \frac{1}{v_0^d(t)} G(\vec{c}),
\end{equation}
where the thermal velocity $v_0^2(t)$ is given by $v_0^2(t) = \frac{2 \langle \vec{v}^2 \rangle}{d}$, and the function $G(\vec{c})$ is expanded in terms of Sonine polynomials as:
\begin{equation}
\label{sonine_expansion}
G(\vec{c}) = \frac{1}{\pi^{d/2}} \exp(-c^2) \sum_{n=0}^{\infty} a_n S_n(c^2).
\end{equation}
In the above equation (\ref{sonine_expansion}), the velocity distribution $G(\vec{c})$ has been represented using the Sonine polynomial series. For reference, the first few Sonine polynomials are as follows:
\begin{align*}
S_0(c^2) &= 1, \\
S_1(c^2) &= \frac{d}{2} - c^2, \\
S_2(c^2) &= \frac{d(d+2)}{8} - \frac{(d+2)}{2} c^2 + \frac{c^4}{2}, \\
S_3(c^2) &= \frac{d(d+2)(d+4)}{48} - \frac{(d+2)(d+4)}{8} c^2 + \frac{(d+4)}{4} c^4 - \frac{c^6}{6}, \\
S_4(c^2) &= \frac{d(d+2)(d+4)(d+6)}{384} - \frac{(d+2)(d+4)(d+6)}{48} c^2 \\
&\quad + \frac{(d+4)(d+6)}{16} c^4 - \frac{(d+6)}{12} c^6 + \frac{c^8}{24}, \\
S_5(c^2) &= \frac{d(d+2)(d+4)(d+6)(d+8)}{3840} \\
&\quad - \frac{(d+2)(d+4)(d+6)(d+8)}{384} c^2 \\
&\quad + \frac{(d+4)(d+6)(d+8)}{96} c^4 \\
&\quad - \frac{(d+6)(d+8)}{48} c^6 \\
&\quad + \frac{(d+8)}{48} c^8 \\
&\quad - \frac{c^{10}}{120}.
\end{align*}
These polynomials satisfy the following orthogonality condition:
\begin{equation}
\int_0^{\infty} dc \, c^{d-1} \exp(-c^2) S_n(c^2) S_m(c^2) = \delta_{nm} \frac{\Gamma(n + d/2)}{2 n!}.
\end{equation}
The deviation from the Maxwell-Boltzmann velocity distribution can be measured through the coefficients $a_n$ in the Sonine polynomial expansion. In the case of no dissipation, all coefficients except the leading one vanish. However, when dissipation is introduced, all coefficients $a_n$ (for $n \geq 2$) take non-zero values, with $a_0 = 1$ and $a_1 = 0$ in both scenarios.
By employing the methods of kinetic theory, Brilliantov and Pöschel (BP) derived the expressions for the first two nontrivial Sonine coefficients ($a_2$ and $a_3$) in the homogeneous cooling state (HCS). For $d = 3$, these expressions are given as follows:
 \cite{bpepl2006}:
\begin{align}
    a_{2} &= -\frac{16}{c(e)}\left(-1623 + 1934 e + 895 e^{2} - 364 e^{3} \right. \nonumber \\
    & \quad \left. + 3510 e^{4} - 7424 e^{5} + 3312 e^{6} - 480 e^{7} + 240 e^{8}\right), \nonumber \\
    a_{3} &= -\frac{128}{c(e)}\left(217 - 386 e - 669 e^{2} + 1548 e^{3} + 154 e^{4} \right. \nonumber \\
    & \quad \left. - 1600 e^{5} + 816 e^{6} - 160 e^{7} + 80 e^{8}\right), \nonumber \\
    c(e) &= 214357 - 172458 e + 112155 e^{2} + 25716 e^{3} - 4410 e^{4} \nonumber \\
    & \quad - 84480 e^{5} + 34800 e^{6} - 5600 e^{7} + 2800 e^{8}
\end{align}
In order to obtain the time evolution of $a_n(t)$, we use the expansion 
\begin{equation}
    \langle c^{2k}\rangle(t) = \langle c^{2k}\rangle_{\text{MB}} \sum_{n=0}^{k}(-1)^n \frac{k!}{n!(k-n)!} a_n(t)
\end{equation}
where 
\begin{equation}
    \langle c^{2k}\rangle_{\text{MB}} = \frac{\Gamma(k+d/2)}{\Gamma(d/2)}.
\end{equation}
This yields the first few \( a_n \)'s as follows:
\begin{align}
    a_1(t) &= 1 - \frac{\left\langle c^2\right\rangle}{\left\langle c^2\right\rangle_{\mathrm{MB}}} = 0, \nonumber \\ 
    a_2(t) &= -1 + \frac{\left\langle c^4\right\rangle}{\left\langle c^4\right\rangle_{\mathrm{MB}}}, \nonumber \\
    a_3(t) &= 1 + 3 a_2 - \frac{\left\langle c^6\right\rangle}{\left\langle c^6\right\rangle_{\mathrm{MB}}}, \nonumber \\
    a_4(t) &= -1 - 6 a_2 + 4 a_3 + \frac{\left\langle c^8\right\rangle}{\left\langle c^8\right\rangle_{\mathrm{MB}}}, \nonumber \\
    a_5(t) &= 1 + 10 a_2 - 10 a_3 + 5 a_4 - \frac{\left\langle c^{10}\right\rangle}{\left\langle c^{10}\right\rangle_{\mathrm{MB}}}, \text{ etc. } \nonumber
\end{align}
Figure \ref{fig:residual_z} shows the residuals for the X-direction velocity distribution.

\begin{figure}[htbp]
    \centering
    \includegraphics[width=0.50\textwidth]{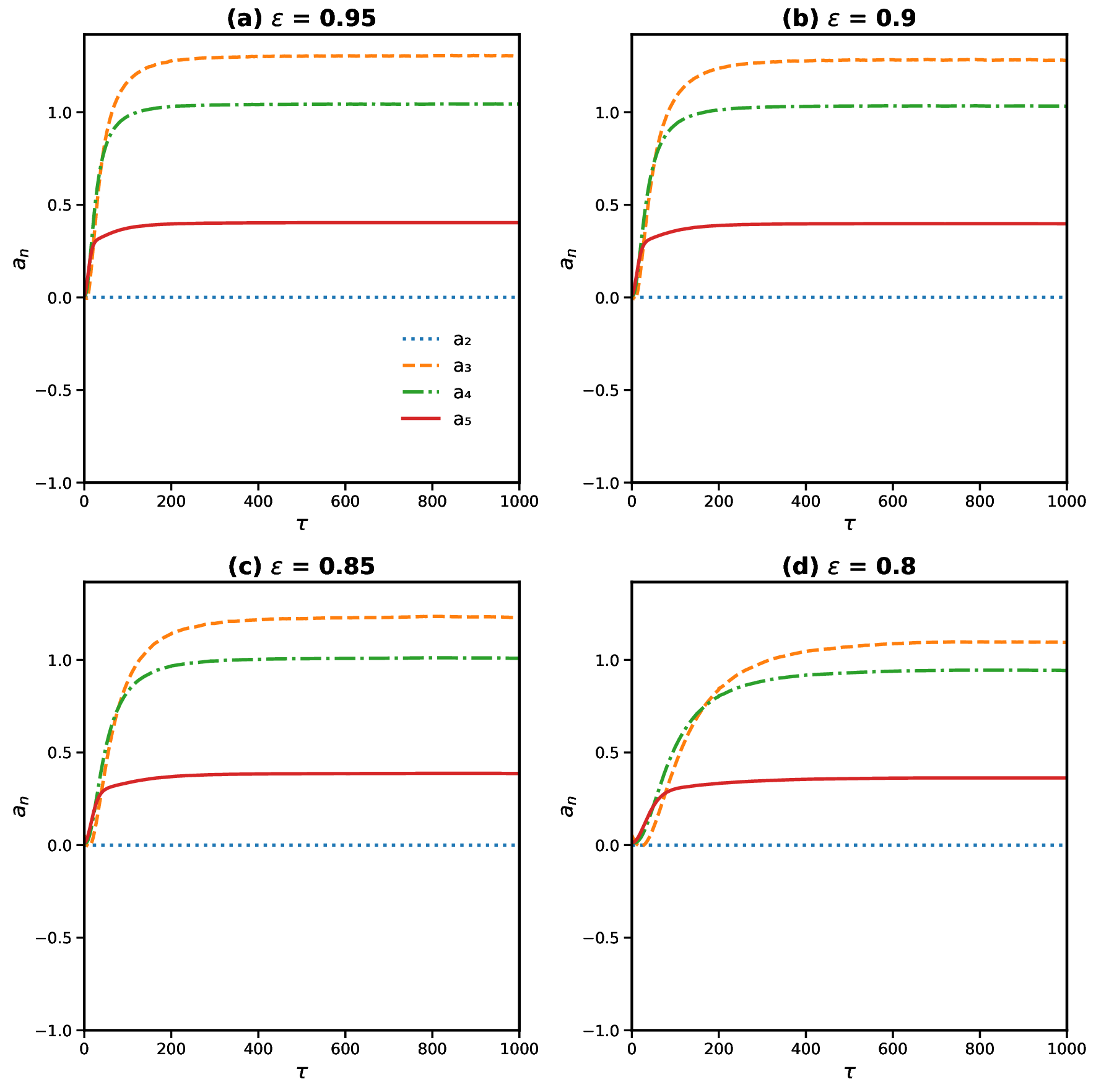} 
    \caption{ Time evolution of Sonine coefficients $a_2$, $a_3$, $a_4$, and $a_5$ for a vibrated binary granular gas system with different coefficients of restitution: (a) $\varepsilon = 0.95$, (b) $\varepsilon = 0.90$, (c) $\varepsilon = 0.85$, and (d) $\varepsilon = 0.80$. The x-axis represents dimensionless time $\tau$, and the y-axis shows the values of the Sonine coefficients $a_n$.}
    \label{f3sonine}
\end{figure}

The Maxwell-Boltzmann (MB) distribution provides a foundational understanding of how particle velocities are distributed in equilibrium gases. In our non-equilibrium system of inelastic particles subjected to vibration, we expect—and indeed observe—significant deviations from the MB distribution. To quantify these deviations, we conducted a detailed analysis of the velocity distributions in each direction (X, Y, and Z) for various restitution coefficients, specifically $\epsilon = 0.80, 0.85, 0.90, \text{ and } 0.95$.
For each direction, we plotted the simulated velocity distribution against the MB distribution. Notable deviations from the MB distribution are evident, particularly in the tails of the distribution.
To further elucidate these deviations, we calculated the residuals (the difference from the MB distribution) and the ratio of our observed distribution to the MB distribution. The ratio plot clearly indicates that our distribution exhibits heavier tails compared to the MB distribution, with the ratio exceeding 1 for larger velocity magnitudes. This observation suggests an increased probability of finding particles with velocities significantly larger than the mean in our vibrated granular system.
These findings underscore the non-equilibrium nature of our system and highlight the importance of understanding velocity distribution anomalies in driven granular materials.
The residuals reveal systematic deviations from the MB distribution, with positive residuals in the tails indicating an excess of high-velocity particles compared to the equilibrium expectation. The ratio plot clearly shows that our distribution has heavier tails than the MB distribution, with the ratio exceeding 1 for larger velocity magnitudes. This indicates an increased probability of finding particles with velocities much larger than the mean in our vibrated granular system.
Next, we investigate the time evolution of the coefficients in the Sonine polynomial expansion. The departure from the Maxwell-Boltzmann velocity distribution function (VDF) is indicated by the non-zero values of the coefficients $a_k$, where $k \geq 2$. In Fig. \ref{f3sonine}, we plot the Sonine coefficients $a_2, a_3, a_4,$ and $a_5$ as a function of $\tau$ for (a) $\epsilon = 0.95$, (b) $\epsilon = 0.90$, (c) $\epsilon = 0.85$, and (d) $\epsilon = 0.80$. It is evident that the Sonine coefficients for all values of $\epsilon$ converge to non-zero values. Higher-order coefficients are observed to approach progressively smaller values, confirming the convergence of the series expansion.
Figure~\ref{f3sonine} illustrates the time evolution of the Sonine polynomial coefficients ($a_2$, $a_3$, $a_4$, and $a_5$) for a vibrated binary granular gas system with varying coefficients of restitution ($\varepsilon$). This figure provides crucial insights into the non-equilibrium behavior of the system and its deviation from the Maxwell-Boltzmann distribution. The key observations from this figure are as follows:
 For all values of $\varepsilon$, the Sonine coefficients converge to non-zero values over time. This convergence indicates that the system reaches a steady state that significantly deviates from the Maxwell-Boltzmann distribution, which would be characterized by all coefficients being zero except $a_0 = 1$ and $a_1 = 0$.
In all cases, we observe $|a_3| > |a_4| > |a_5| > |a_2|$. This decreasing trend in magnitude for successive coefficients implies the convergence of the Sonine polynomial expansion and demonstrates the relative importance of each term in characterizing the velocity distribution.
  As $\varepsilon$ decreases (i.e., collisions become more inelastic), the magnitudes of the Sonine coefficients generally increase. This trend suggests that more inelastic collisions lead to greater deviations from the Maxwell-Boltzmann distribution, reflecting the enhanced non-equilibrium nature of the system.
After an initial transient period ($\tau < 200$), the coefficients stabilize, indicating that the system reaches a non-equilibrium steady state. This stability is maintained even for highly inelastic collisions ($\varepsilon = 0.80$), demonstrating the robustness of the steady state.
 The non-zero values of these coefficients, particularly for $n \geq 2$, provide clear evidence for the non-equilibrium nature of the vibrated granular system. In an equilibrium system, these coefficients would be zero, as the velocity distribution would follow the Maxwell-Boltzmann form.
\subsection{Statistical Moments}
To quantify the shape of our velocity distributions, we calculated the kurtosis and skewness for each direction. These statistical measures provide insight into the distribution's tails and asymmetry, which are crucial for understanding the non-equilibrium dynamics in our vibrated binary granular gas system [TABLE I].
\begin{figure*}[htbp] 
    \centering
    
    \begin{minipage}[b]{0.45\textwidth}
        \centering
        \includegraphics[width=\textwidth]{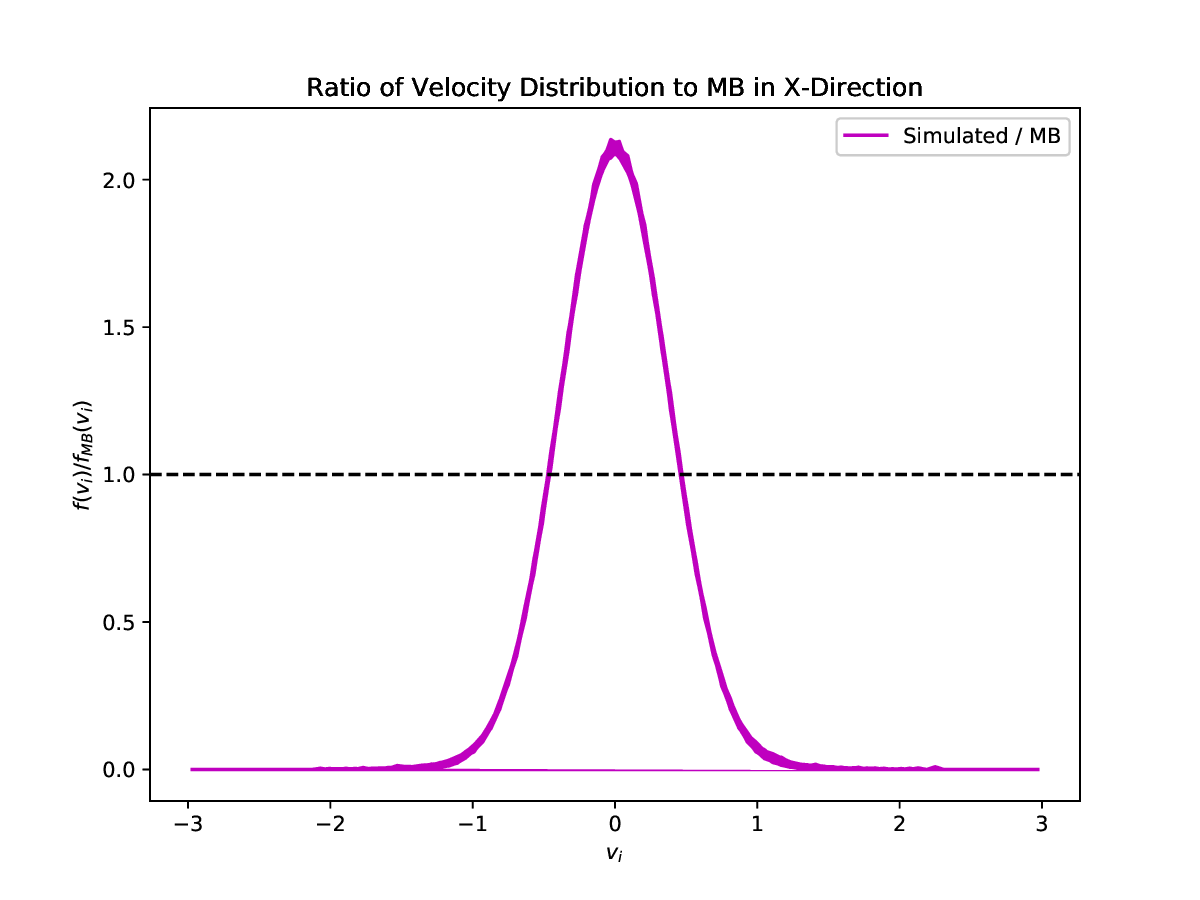}
        \caption*{Ratio of simulated velocity distribution to Maxwell-Boltzmann distribution in the X-direction for restitution coefficient $\epsilon = 0.80$.}
        \label{fig:velocity_distribution_0.80}
    \end{minipage}
    \hfill
    \begin{minipage}[b]{0.45\textwidth}
        \centering
        \includegraphics[width=\textwidth]{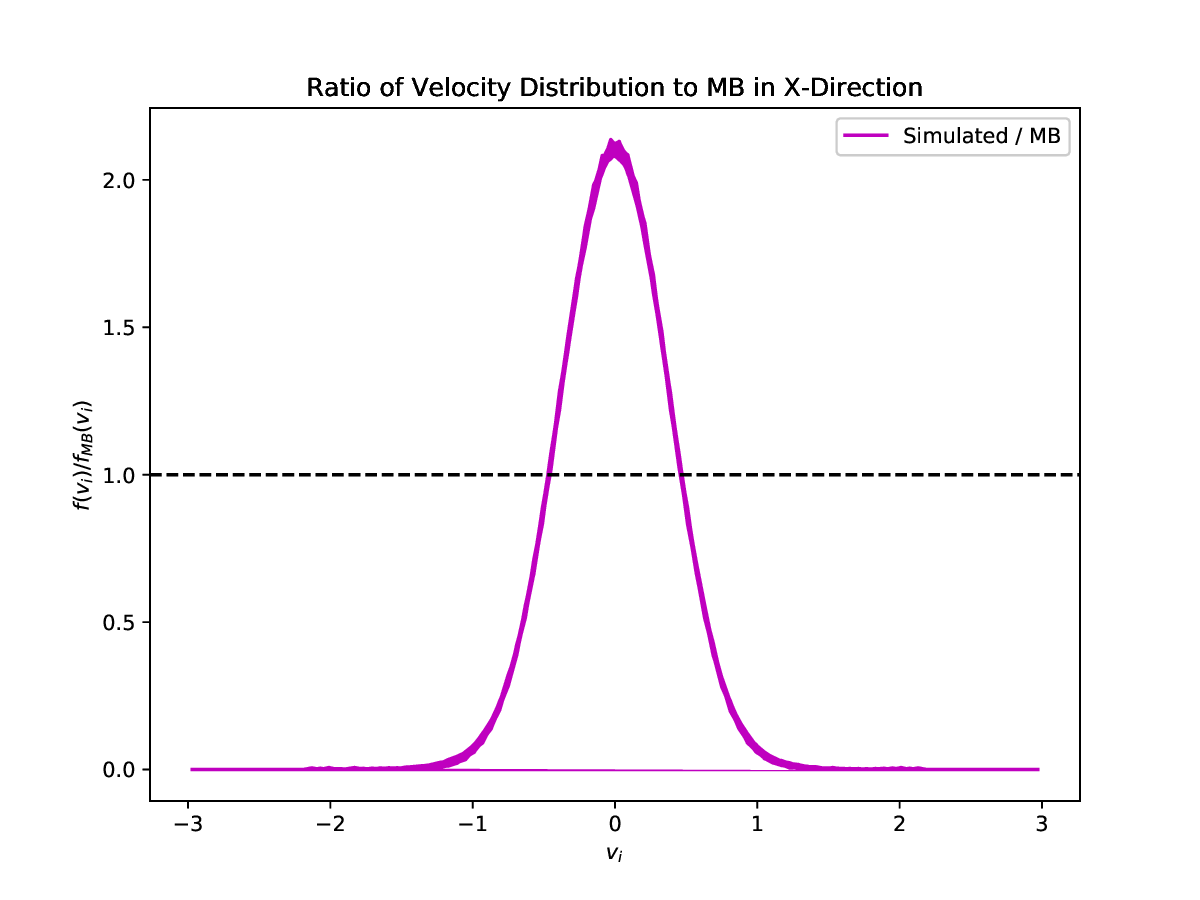}
        \caption*{Ratio of simulated velocity distribution to Maxwell-Boltzmann distribution in the X-direction for restitution coefficient $\epsilon = 0.85$.}
        \label{fig:ratio_0.85_z}
    \end{minipage}

    \vskip\baselineskip
    
    \begin{minipage}[b]{0.45\textwidth}
        \centering
        \includegraphics[width=\textwidth]{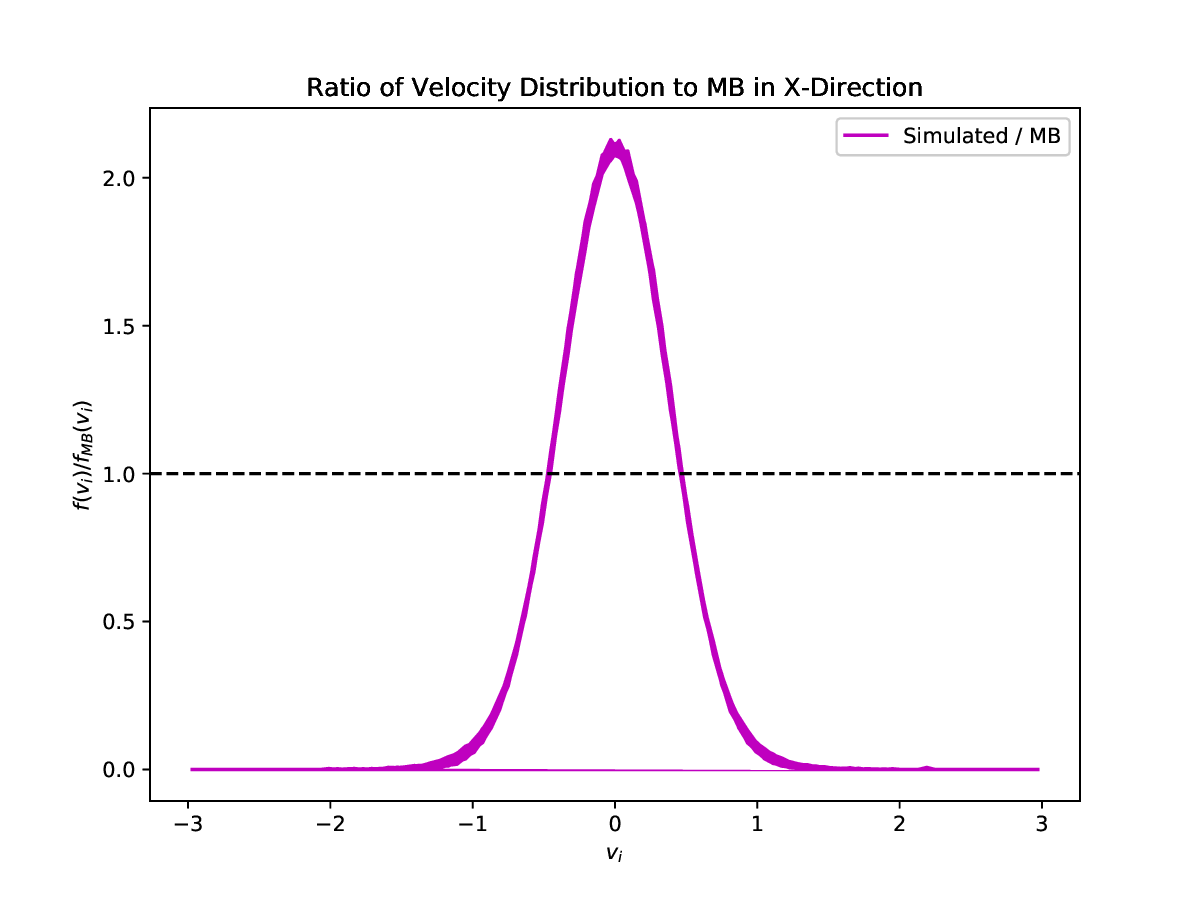}
        \caption*{Ratio of simulated velocity distribution to Maxwell-Boltzmann distribution in the X-direction for restitution coefficient $\epsilon = 0.90$.}
        \label{fig:ratio_0.90_z}
    \end{minipage}
    \hfill
    \begin{minipage}[b]{0.45\textwidth}
        \centering
        \includegraphics[width=\textwidth]{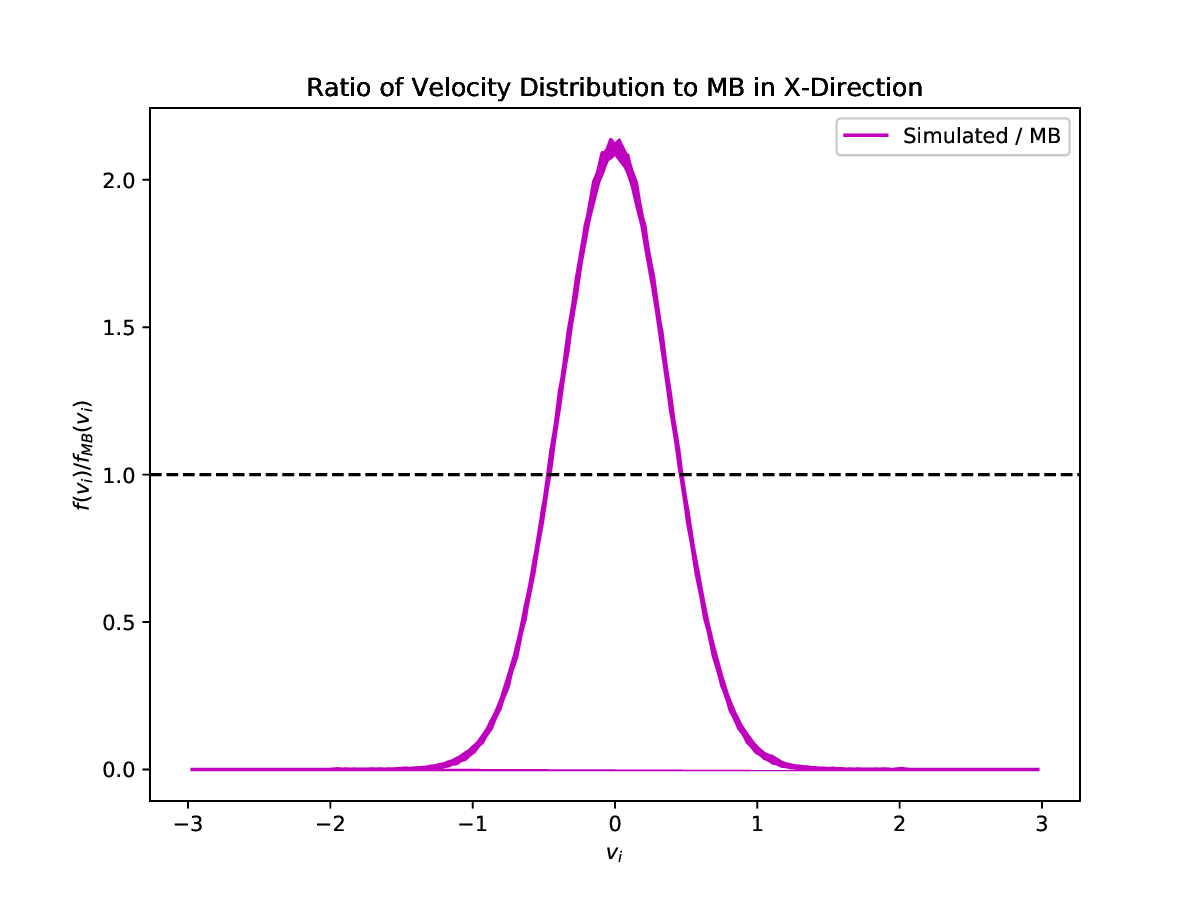}
        \caption*{Ratio of simulated velocity distribution to Maxwell-Boltzmann distribution in the X-direction for restitution coefficient $\epsilon = 0.95$.}
        \label{fig:ratio_0.95_z}
    \end{minipage}

    \caption{ The ratio plots clearly show that our distribution has heavier tails than the Maxwell-Boltzmann (MB) distribution, with the ratio exceeding 1 for larger velocity magnitudes. This indicates an increased probability of finding particles with velocities much larger than the mean in our vibrated granular system.}
    \label{fig:ratio_velocity_distribution_z}
\end{figure*}
Figure~\ref{fig:velocity_distribution} shows the velocity distribution \( v_0 P(v, \tau) \) as a function of the normalized velocity \( v/v_0 \) for a binary granular gas system with a coefficient of restitution \( \epsilon = 0.80, 0.85, 0.90 \, \& \, 0.95 \). The distribution highlights the non-equilibrium nature of the system due to the inelastic collisions between particles. The red data points represent simulation results, while the blue line shows a smooth fit to the data. The plot reveals that the system exhibits a peaked velocity distribution, with the most probable velocities occurring around \( v/v_0 \approx 0.5 \), followed by a rapid decay at higher velocities. The presence of a "tail" in the distribution indicates that there are still particles with high velocities, albeit with a lower probability. The deviation from the classical Maxwell-Boltzmann distribution is evident, showcasing the effects of energy dissipation in granular systems driven by external forces such as vibration.
This distribution reflects the system's far-from-equilibrium state, where inelastic collisions lead to a clustering of velocities around lower values, resulting in non-Maxwellian behavior. Such findings are crucial for understanding how granular gases behave under various conditions, especially in terms of their energy distribution and transport properties.
\begin{figure*}[htbp] 
    \centering
    
    \begin{minipage}[b]{0.45\textwidth}
        \centering
        \includegraphics[width=\textwidth]{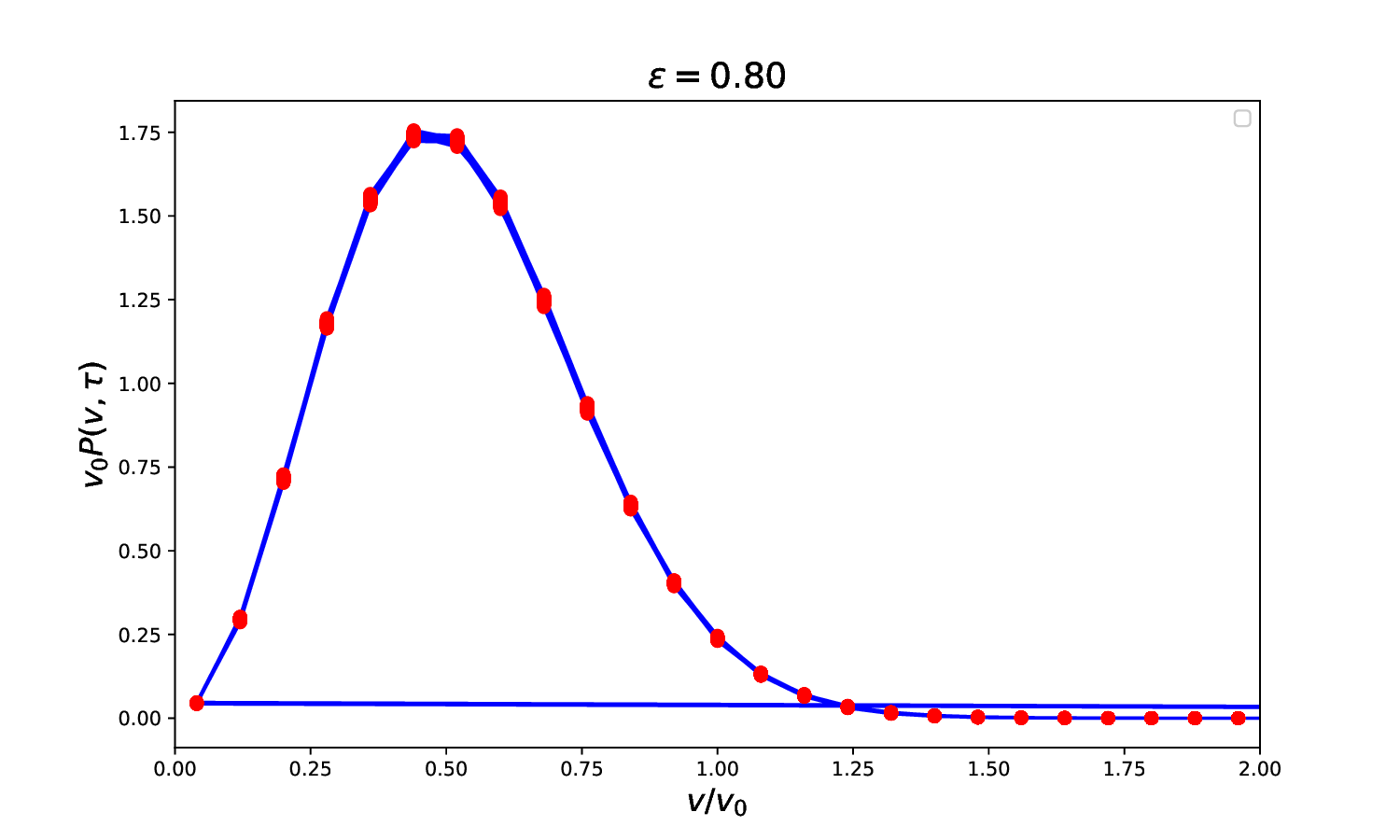}
        \caption*{Velocity distribution \( v_0 P(v, \tau) \) for restitution coefficient \( \epsilon = 0.80 \).}
        \label{fig:velocity_distribution_0.80}
    \end{minipage}
    \hfill
    \begin{minipage}[b]{0.45\textwidth}
        \centering
        \includegraphics[width=\textwidth]{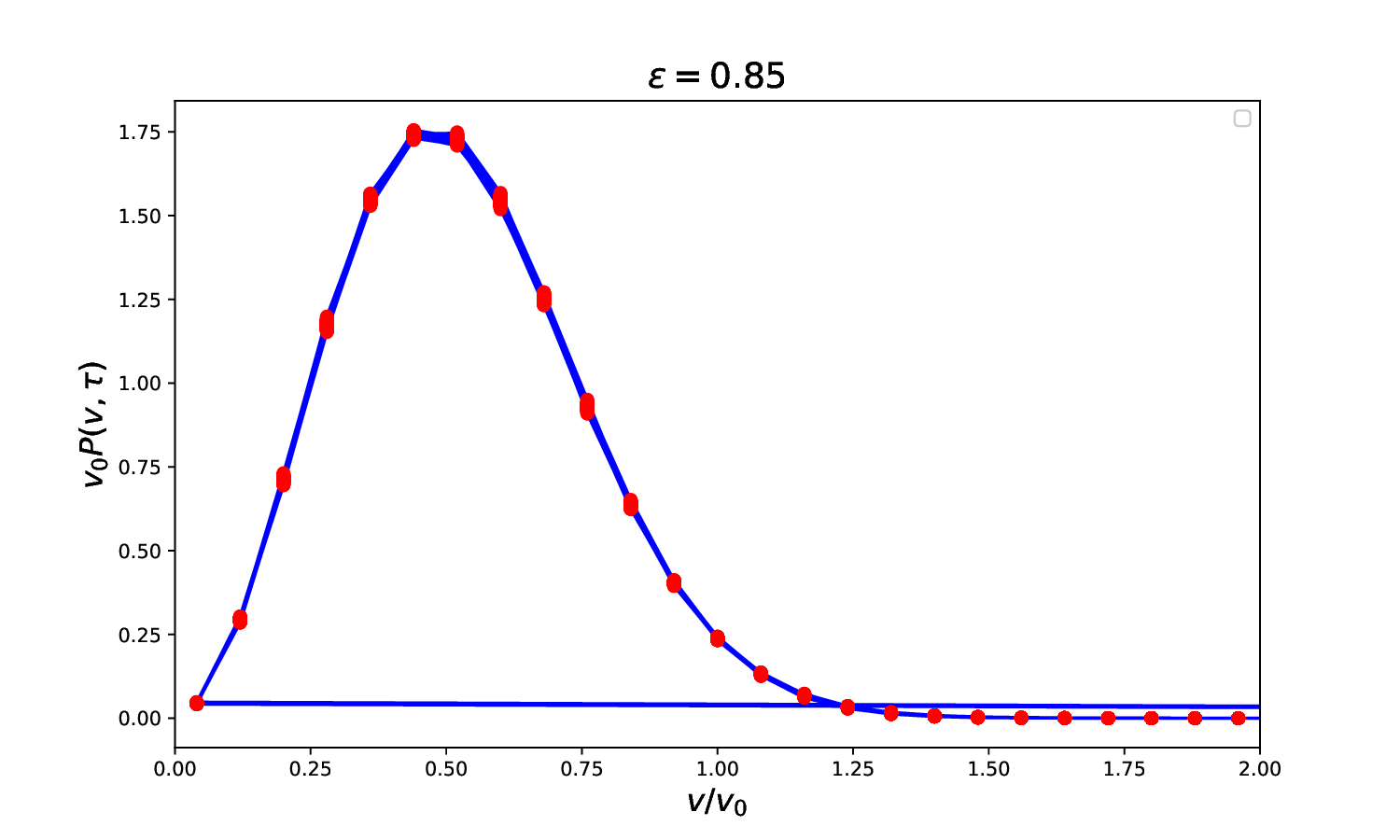}
        \caption*{Velocity distribution \( v_0 P(v, \tau) \) for restitution coefficient \( \epsilon = 0.85 \).}
        \label{fig:ratio_0.85_z}
    \end{minipage}

    \vskip\baselineskip
    
    \begin{minipage}[b]{0.45\textwidth}
        \centering
        \includegraphics[width=\textwidth]{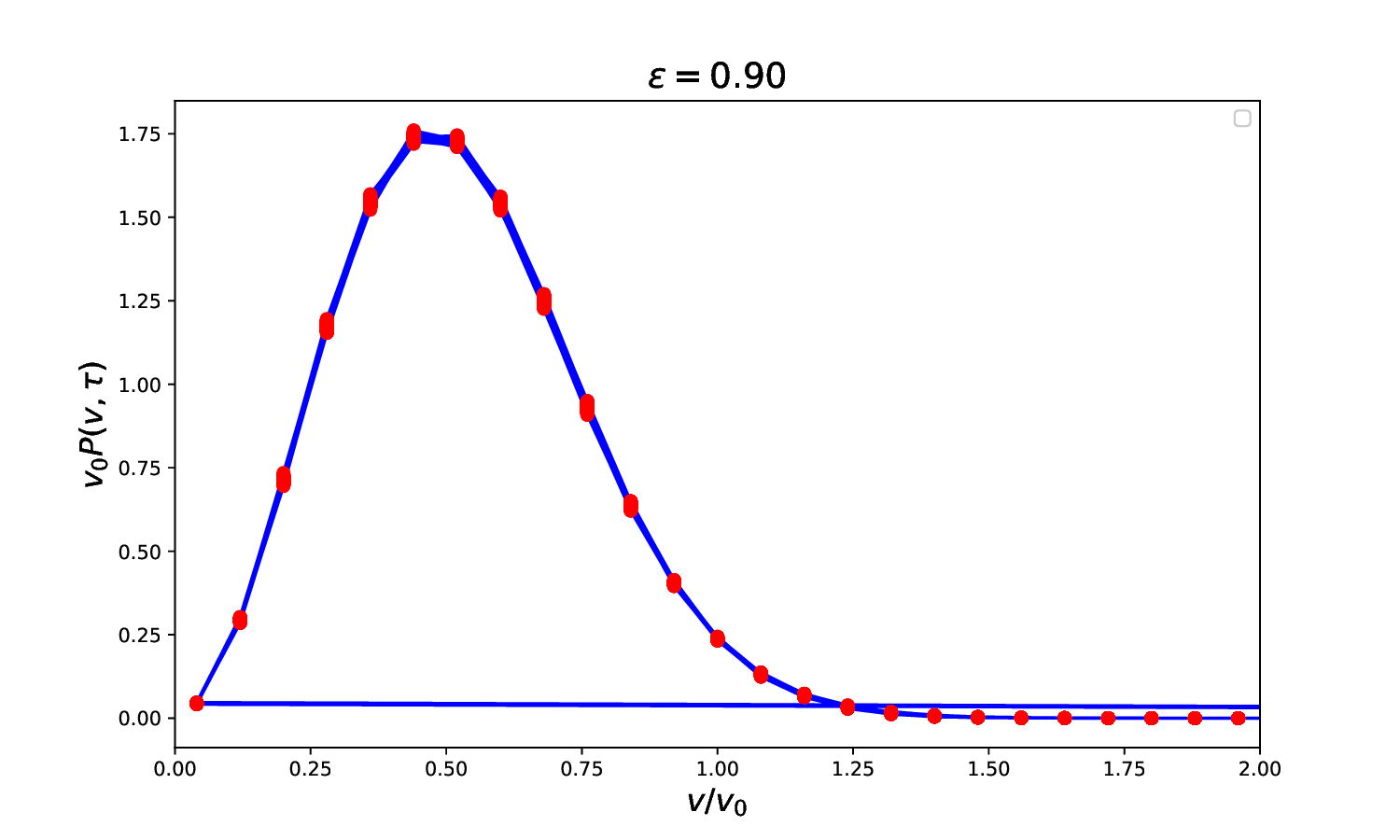}
        \caption*{Velocity distribution \( v_0 P(v, \tau) \) for restitution coefficient \( \epsilon = 0.90 \).}
        \label{fig:ratio_0.90_z}
    \end{minipage}
    \hfill
    \begin{minipage}[b]{0.45\textwidth}
        \centering
        \includegraphics[width=\textwidth]{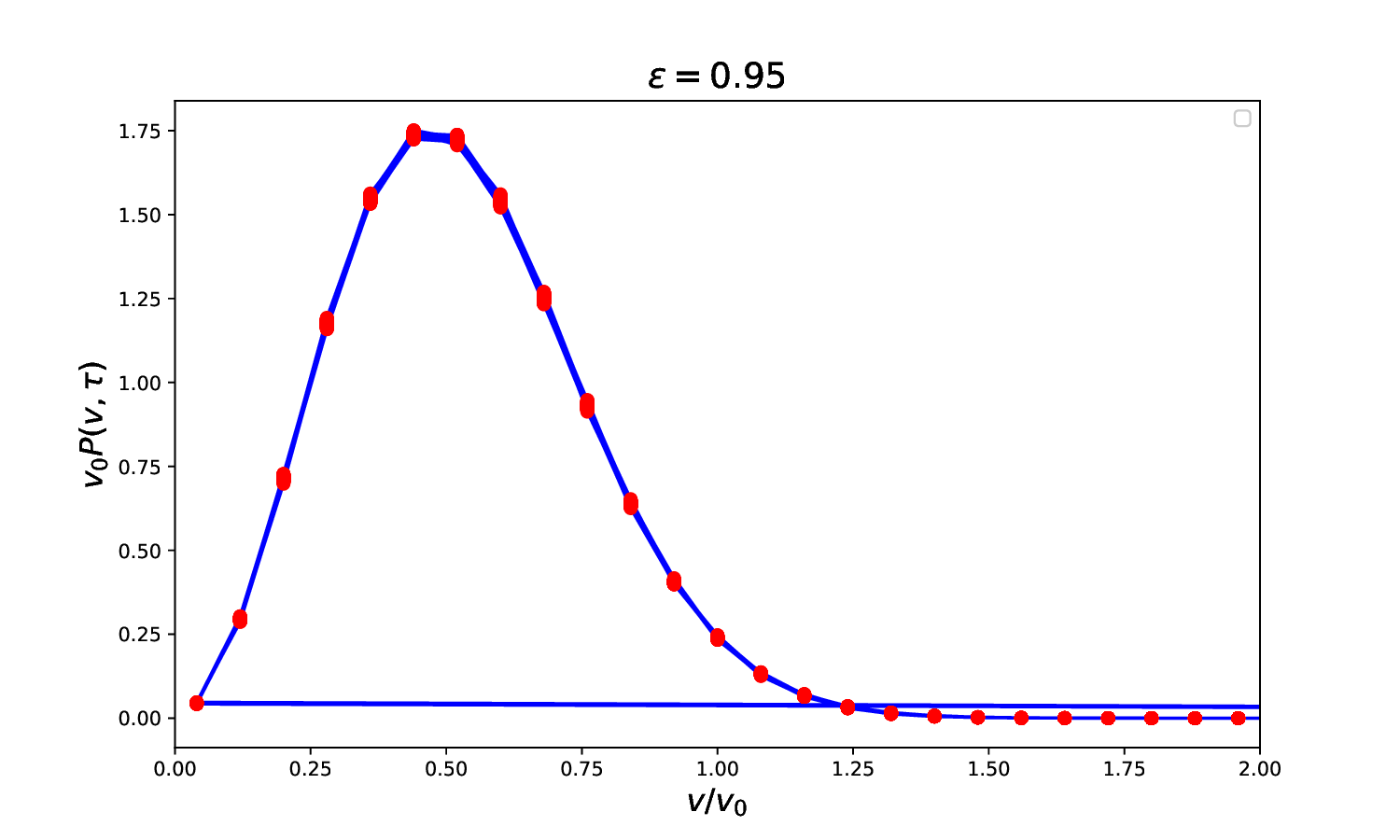}
        \caption*{Velocity distribution \( v_0 P(v, \tau) \) for restitution coefficient \( \epsilon = 0.95 \).}
        \label{fig:ratio_0.95_z}
    \end{minipage}

    \caption{FIG. 3: Velocity distribution \( v_0 P(v, \tau) \) as a function of normalized velocity \( v / v_0 \) for a binary granular gas system with coefficients of restitution \( \epsilon = 0.80, 0.85, 0.90, \text{ and } 0.95 \). The red points represent simulation data, and the blue line is a smooth fit to the distribution. The distribution exhibits a peaked profile around \( v / v_0 \approx 0.5 \) and a heavy tail, indicating a higher probability of lower velocities due to inelastic collisions. This behavior deviates from the classical Maxwell-Boltzmann distribution, characteristic of granular systems with energy dissipation.}
    \label{fig:velocity_distribution}
\end{figure*}
\begin{table*}[htbp] 
\centering
\caption{Kurtosis and Skewness of Velocity Distribution in the X-Direction for Different Coefficients of Restitution $(\epsilon)$}
\begin{tabular*}{\textwidth}{@{\extracolsep{\fill}} D{.}{.}{8} D{.}{.}{18} D{.}{.}{18} @{}}
\toprule
\textbf{Coefficient of Restitution $(\epsilon)$} & \textbf{Kurtosis} & \textbf{Skewness} \\
\midrule
0.80 & 2.59227390 & 2.00126436 \\
0.85 & 2.58489102 & 1.99957497 \\
0.90 & 2.56248216 & 1.99452220 \\
0.95 & 2.58775779 & 2.00028713 \\
\bottomrule
\end{tabular*}
\end{table*}
The positive kurtosis values (approximately 2.58–2.59) indicate that our velocity distribution exhibits heavier tails compared to a Gaussian distribution, suggesting an increased probability of observing extreme velocities. This finding aligns with our observations from the ratio plot, where we noted a higher frequency of particles with velocities significantly greater than the mean.
Similarly, the skewness values (around 1.99–2.00) reveal a rightward asymmetry in the distribution, indicating that there are more high-velocity particles than low-velocity ones. This asymmetry is characteristic of driven systems, where energy input through vibration leads to a concentration of particles with high kinetic energy. Collectively, these results provide strong evidence for the non-equilibrium nature of our vibrated binary granular gas system. The enhanced high-energy tails in the velocity distribution reflect the interplay between energy input from vibration and energy dissipation through inelastic collisions. Specifically, the vibration mechanism is crucial for populating the high-velocity tails, resulting in a distribution that significantly deviates from equilibrium expectations.
In the following sections, we will discuss these findings in more detail, exploring how they vary with the coefficient of restitution and vibration intensity, as well as examining their implications for the statistical mechanics of driven granular systems.
\section{\label{sec:summry} SUMMARY AND CONCLUSION}
In this study, we conducted a detailed analysis of velocity distributions in a vibrated binary granular gas system using molecular dynamics simulations. Our focus was on understanding how inelastic collisions, governed by the coefficient of restitution (CoR), affect the system's velocity statistics. The results clearly show that as the CoR decreases, the system deviates further from the classical Maxwell-Boltzmann (MB) distribution. The presence of heavier tails in the velocity distributions indicates a higher likelihood of particles having extreme velocities, which is characteristic of non-equilibrium steady states driven by external energy input.
The deviations from the MB distribution were further quantified through statistical moments, such as kurtosis and skewness. These metrics confirm that the system's velocity distribution exhibits both heavy tails and a rightward asymmetry, reflecting the non-equipartition of energy between different particle types. Notably, this asymmetry becomes more pronounced at lower CoR values, emphasizing the critical role of inelasticity in dictating the dynamics of the system. The observed non-equilibrium behavior, driven by external vibration and dissipative collisions, highlights the complex interplay between energy injection and dissipation mechanisms in granular systems.
Additionally, our results have broader implications for the study of driven granular gases and other non-equilibrium systems. The ability to systematically control factors such as CoR, particle size, and mass ratios provides a powerful tool for tuning the system’s behavior, with potential applications in optimizing industrial processes involving vibrated granular materials, such as mixing, segregation, and transport in bulk solids handling. 
In the steady state, we examined the time evolution of the coefficients in the Sonine polynomial expansion of the velocity distribution function. The deviation from the Maxwell-Boltzmann (MB) distribution is indicated by non-zero values of the Sonine coefficients $a_n$ for $n \geq 2$. In our simulations, the Sonine coefficients $a_2$ through $a_5$ were computed numerically and were found to converge to non-zero values. 
These results underscore the complex dynamics inherent in vibrated granular systems and highlight the significance of inelastic collisions in shaping the velocity distribution. The observed deviations from the Maxwell-Boltzmann distribution, quantified by the Sonine coefficients, reflect the interplay between energy input through vibration and energy dissipation through inelastic collisions. This analysis provides valuable insights into the statistical mechanics of driven granular gases and demonstrates the power of the Sonine polynomial expansion in characterizing non-equilibrium steady states.
Furthermore, the insights gained from this study contribute to the growing body of knowledge in non-equilibrium statistical mechanics and may offer new perspectives on the thermodynamics of granular systems. Future work could explore the effects of varying vibration intensities, particle interactions beyond binary systems, and the impact of confinement on the observed dynamics. These directions would deepen our understanding of the complex behavior in vibrated granular materials and potentially lead to new discoveries in granular physics.
\section*{\label{ack} ACKNOWLEDGEMENTS}
RFS acknowledges financial support from the University Grants Commission in the form of Non-NET fellowships. He also wishes to acknowledge the computational facilities at the Department of Physics, JMI.

\end{document}